\documentclass{amsart}%
\usepackage{amsfonts}
\usepackage{color}
\usepackage{amsmath}
\usepackage{amssymb}
\usepackage{graphicx}
\usepackage{cite}
\providecommand{\U}[1]{\protect\rule{.1in}{.1in}}

\theoremstyle{plain}

\numberwithin{equation}{section}

\begin{document}
\title[A cell-merge algorithm for generating spatial tessellations]{A unified cell-merge algorithm for generating diverse Voronoi diagrams and new
tessellations based on spatial chromatic model}
\author{Weining Zhu}
\address[Zhu, W. N.]{Laboratory of Geospatial Information, Zhejiang University.}
\email{zhuwn@zju.edu.cn}
\date{June 4, 2026}

\begin{abstract}
As a type of spatial tessellation model and an important spatial structure of
computational geometry, Voronoi diagrams (VDs) are widely used in many fields.
Due to differences in generation spaces, types of spatial entities, distance
metrics, and relationships between entities and Voronoi regions, Voronoi
diagrams vary into many types, such as the ordinary VD, VD on spheres, VD for
linear entities, weighted VD, and ordered higher-order VD. These VDs also have
their own generation algorithms. In this study, we propose a new cell-merge
(CM) Voronoi generation algorithm based on the spatial chromatic model. The
advantage of the CM algorithm is that it can quickly generate diverse VDs by
retrieving and merging cells from a unified database, without requiring the development
of specific algorithms for each VD. The CM Voronoi algorithm can be
particularly applied in cases where a variety of Voronoi diagrams are
frequently required for computation and analysis, such as in location-based
spatial analysis. Furthermore, the proposed CM method can also generate some
new types of spatial tessellations, such as competition intensity and couple-cell diagrams which are different from classical VDs.

\end{abstract}
\maketitle

\section{Introduction}

\setlength{\parskip}{3pt}The Voronoi diagram (VD), also known as a spatial
tessellation or partition model \cite{Aurenhammer1991,Okabe2000}, was originally theorized and
studied by the mathematicians Descartes (French), Dirichlet (German), and
Voronoi (Ukrainian, after whom it is named). Current VD research spans
mathematics, computer science (computational geometry, graphics, and
visualization), and applications in spatial interpolation, point pattern
analysis, location optimization, geospatial sciences, meteorology, hydrology,
fluid dynamics, materials science, and biology \cite{Gold1994,Chen2001,ORourke1998}. In meteorology,
for example, VDs are known as Thiessen polygons \cite{Thiessen1911} and define the
spatial coverage of weather station observations.

Given a point set $\mathbf{P}=\{p_{1},p_{2},\ldots,p_{n}\}$ in a plane
$\Re^{2}$, where $2<n<\infty$, the Voronoi region $v(p_{i})$ of $p_{i}$ is
defined as:%
\begin{equation}
v(p_{i})=\{p|d(p,p_{i})\leq d(p,p_{j}),\forall j\neq i,j\in\mathbf{I}%
_{n}\},\label{Eq1}%
\end{equation}
where $d$ is a distance metric, and $\mathbf{I}_n = \{1,\dots,n\}$ is the index set of points $p$. The ordinary Voronoi diagram (OVD) of
$\mathbf{P}$ is the union of the Voronoi regions of all points in $\mathbf{P}%
$, i.e., $\mathbf{V}=\{v(p_{1}),v(p_{2}),\ldots,v(p_{n})\}$.

In extensive theoretical studies and practical applications of VDs, numerous
VDs other than the OVD have been proposed, differing in four aspects: (1) Spatial
dimensionality and scope: beyond the most common 2D planar space, some VDs are
constructed in 3D or higher-dimensional spaces, on spherical or conical
surfaces, and in network spaces with paths or nodes, which may also contain
constraints or obstacles. (2) The types of objects in $\mathbf{P}$: the most
common type of object is spatial points, but objects can also be lines,
polygons, or shapes with greater complexity. (3) The distance metric $d$: the
most common metric used in OVDs is the Euclidean distance, but alternatives
include weighted, Manhattan, Hausdorff, and convex distances. (4) Space-object
relationships in Voronoi regions: in the OVD, the relationship is based on the
nearest proximity, whereas in furthest-point VDs, the relationship is changed
to the furthest. Similarly, other relationships generate the higher-order,
ordered, or $k$-th nearest neighbor VDs \cite{Boots1980,Stifter1995,Chew1993,Hanson1983,Hakimi1992}.
Traditional algorithms for generating OVDs and other VDs include incremental,
divide-and-conquer, and plane-sweep algorithms, as well as other specialized
algorithms for high-dimensional, $k$-th nearest neighbor, ordered, obstacle,
linear object, and areal object VDs \cite{Aurenhammer1984,Guibas1992,Amato1996,Fortune1987}. These algorithms vary in
computational cost, complexity, storage space, and applicability. Although
these known methods may work for specific applications, if an application
requires extensive, frequent, and diverse types of Voronoi-based analyses, it
is subject to high development and maintenance costs. To address this
challenge, this study presents a cell-merge algorithm (hereafter referred to
as the CM algorithm) based on the spatial chromatic model (SCM). The CM
algorithm first generates a database of \textquotedblleft
cells\textquotedblright, and then based on the \textquotedblleft genetic
code\textquotedblright\ of these cells, Voronoi regions come out by merging
cells into \textquotedblleft tissues\textquotedblright\ or \textquotedblleft
organs\textquotedblright. To generate different VDs, cells with the same
properties are retrieved from the cell database and then merged. Compared to
traditional solutions, the CM algorithm eliminates the need to set the
specific generation algorithm for each type of VD. Instead, users can quickly
generate their desired VDs by simply modifying a few search parameters in the CM
algorithm. In other words, one algorithm (the CM algorithm) is used to replace
all other VD algorithms. As is well known, database searching is a mature
function that computers perform effectively, and many searching algorithms
have been well developed. For example, Oracle, SQL, and other common database
systems have demonstrated efficient and stable responses in many
massive-data-oriented applications through database design, indexing, data
grouping, and optimized querying \cite{Date2015}. The CM algorithm leverages this
advantage of database searching.

\section{Spatial chromatic model}

\begin{figure}[h]
\centering
\includegraphics[width=1.0\textwidth]{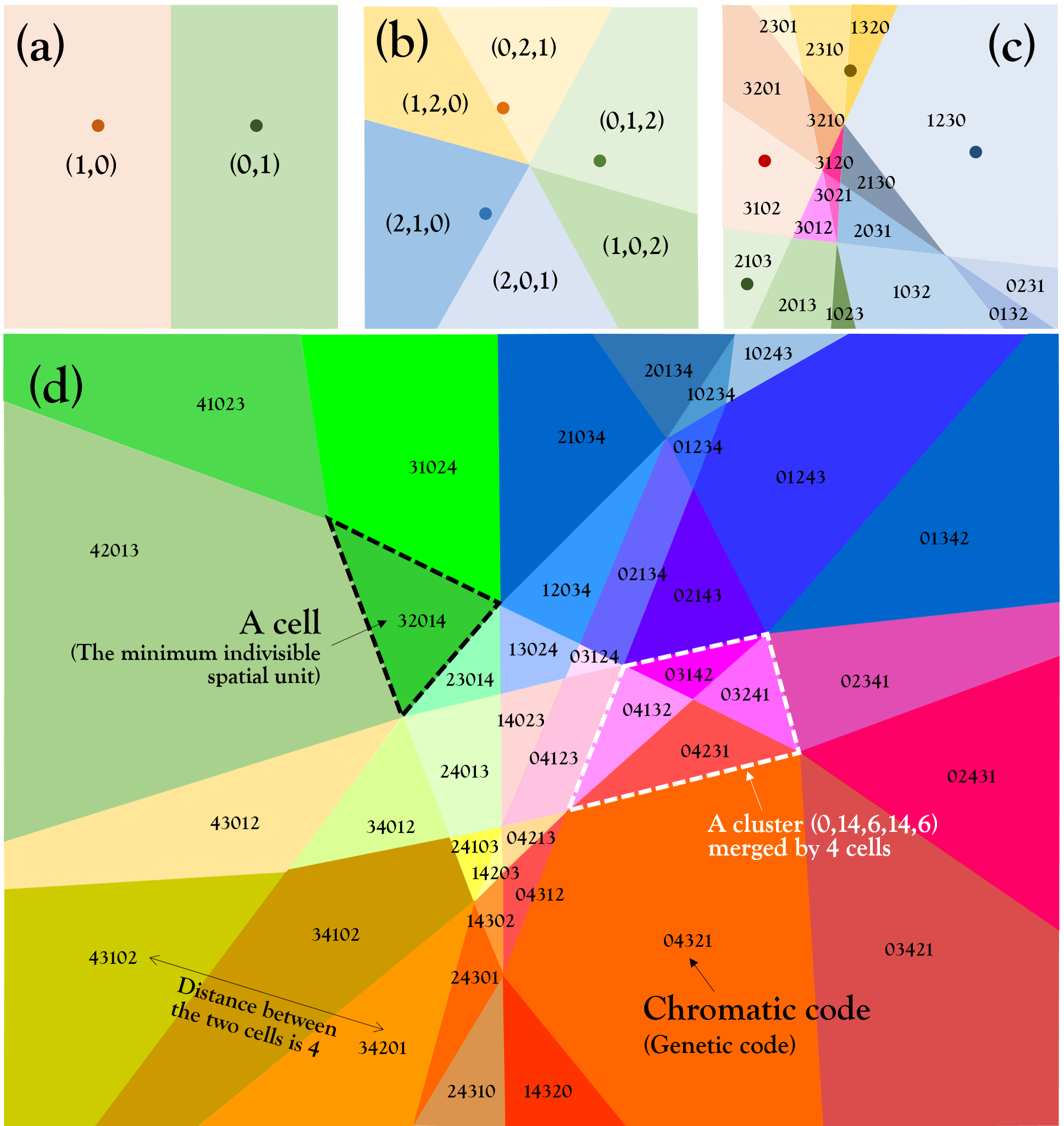}  \caption{SCM space and their
cells generated by (a) two, (b) three, (c) four, and (d) five objects (see
dots in subfigures). Note that the subfigure (d) is a homeomorphous part of
the SCM space generated by five objects, which are not shown in the
subfigure.}%
\label{fig:1}%
\end{figure}

The spatial chromatic model is a recently proposed spatial data model with
applications and implications in spatial topological analysis, spatial
tessellation, spatial pattern recognition, the first law of geography, and
spatial heterogeneity studies \cite{Zhu2010,Zhu2015,Zhu2023}. From the perspective of underlying
spatial data models, SCM is an irregular raster model. Unlike conventional
raster data models, the minimum indivisible spatial unit in SCM is not a
regular pixel or grid but an irregular cell (see Fig. \ref{fig:1}). The spatial
attributes (coordinates) of cells are not represented by row and column
indices $(i,j)$ as in pixels or real number $(x,y)$ in Euclidean plane
$\Re^{2}$, but by chromatic codes that imply spatial relationships. The
chromatic codes of cells are always the arrangement of the integer sequence
$\{0,1,2,...,n\}$, and the integer is the favorite data type which can be
effectively and rapidly processed by computers. The spatial analyses and
operations in SCM, as well as the CM algorithm proposed in this study, are all
based on chromatic codes of cells. Fig. \ref{fig:1} shows the SCMs generated by 2, 3, 4,
and 5 objects, respectively (see Fig. \ref{fig:1}(a)--(d)). Cells in SCM are typically
triangles or polygons with unique chromatic codes. The distance between two
cells can be calculated based on their codes, and multiple cells can merge to
form a larger subspace called a cluster (see an example in Fig. 1(d)), which
also bears the chromatic code inherited from the merged cells.

Generating an SCM space from three objects (see Fig. 1(b)) consists of the
following three steps.

\begin{figure}[h]
\centering
\includegraphics[width=1.0\textwidth]{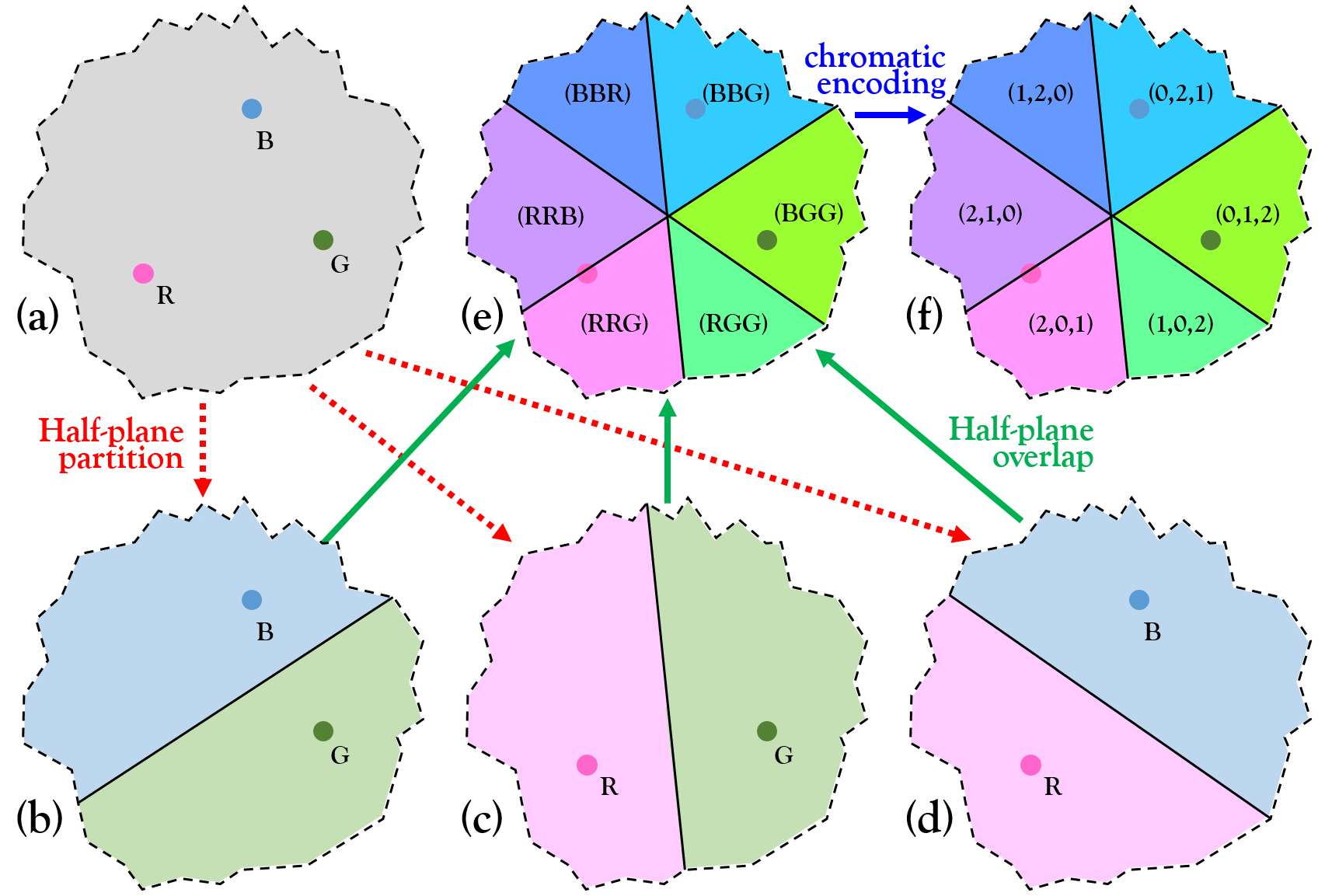}  \caption{An example of the
generating an SCM space by 3 objects. (a) Three objects \{{R, G, B\}} in a 2D
plane; (b), (c), and (d) Half-plane partitioning and dyeing with respect to
the subset family \{{B, G\}}, \{{R, G\}}, and \{{R, B\}}, respectively; (e)
The chromatic space containing 6 cells generated by overlapping the three
spaces in (b), (c) and (d); (f) Chromatic encoding: convert the cell labels
such that (RRB) in (e) to chromatic codes such that (2,1,0) in (f).}%
\label{fig:2}%
\end{figure}

\begin{enumerate}
\item Aggregation. For a set containing three objects $\mathbf{O}=\{$R$,$%
G$,$B$\}$, as shown in Fig. \ref{fig:2}(a), aggregation involves generating a subset
family of the object set. In the standard SCM model, this subset family
consists of all unordered pairs of objects. Thus, for $\mathbf{O}=\{$R$,$%
G$,$B$\}$, the corresponding subset family $\mathbf{F}$ is $\{\{$R$,$%
G$\},\{$G$,$B$\},\{$B$,$R$\}\}$. Each object is assigned a unique color
(equivalent to its index or ID, e.g., R, G, or B).

\item Half-plane partitioning, dyeing, and overlapping. For each element in
the subset family $\mathbf{F}$ (i.e., each pair of objects), the space is
partitioned into two half-planes by a dyeing line. Each half-plane is dyed
with the color of the corresponding object. In the standard SCM model, the
dyeing line is the perpendicular bisector of the line segment connecting the
two objects. After the partitioning and dyeing of all pairs of objects in the
subset family, the space is repeatedly divided into many half-planes (Fig.
\ref{fig:2}(b)--(d)). These half-planes are then overlapped, generating many subspaces
called cells (see six cells in Fig. \ref{fig:2}(e)). Each cell is dyed multiple times by
different object colors during the partitioning processes. Fig. \ref{fig:2} illustrates
the steps of half-plane partitioning, dyeing, and overlapping for three
objects in a 2D plane.

\item Chromatic encoding. To assign chromatic codes to cells, a predefined
object order is first established, e.g., (R, B, G). The chromatic code of a
cell then records the number of times it has been dyed by each object in that
order. For example, a cell with the object label (RRG) as shown in Fig. \ref{fig:2}(e)
is converted to chromatic code (2, 0, 1) in Fig. 2(f), indicating that through
three rounds of partitioning and dyeing, the subspace inside the cell has been
dyed twice by object R, never (zero times) by B, and once by G.
\end{enumerate}

A cell $c$ with a chromatic code in a SCM generated by $n$ objects is usually
denoted by $c(s_{1},s_{2},\ldots,s_{i},\ldots,s_{n})$, in which the integer
$s_{i}$ is called a subcode of the object $o_{i}$ or at the location $i$. The
$n$ subcodes form an arrangement of the integer sequence $(0,1,\ldots,n-1)$.
It is proven that the chromatic code of each cell is unique \cite{Zhu2010} .

\begin{figure}[h]
\centering
\includegraphics[width=1.0\textwidth]{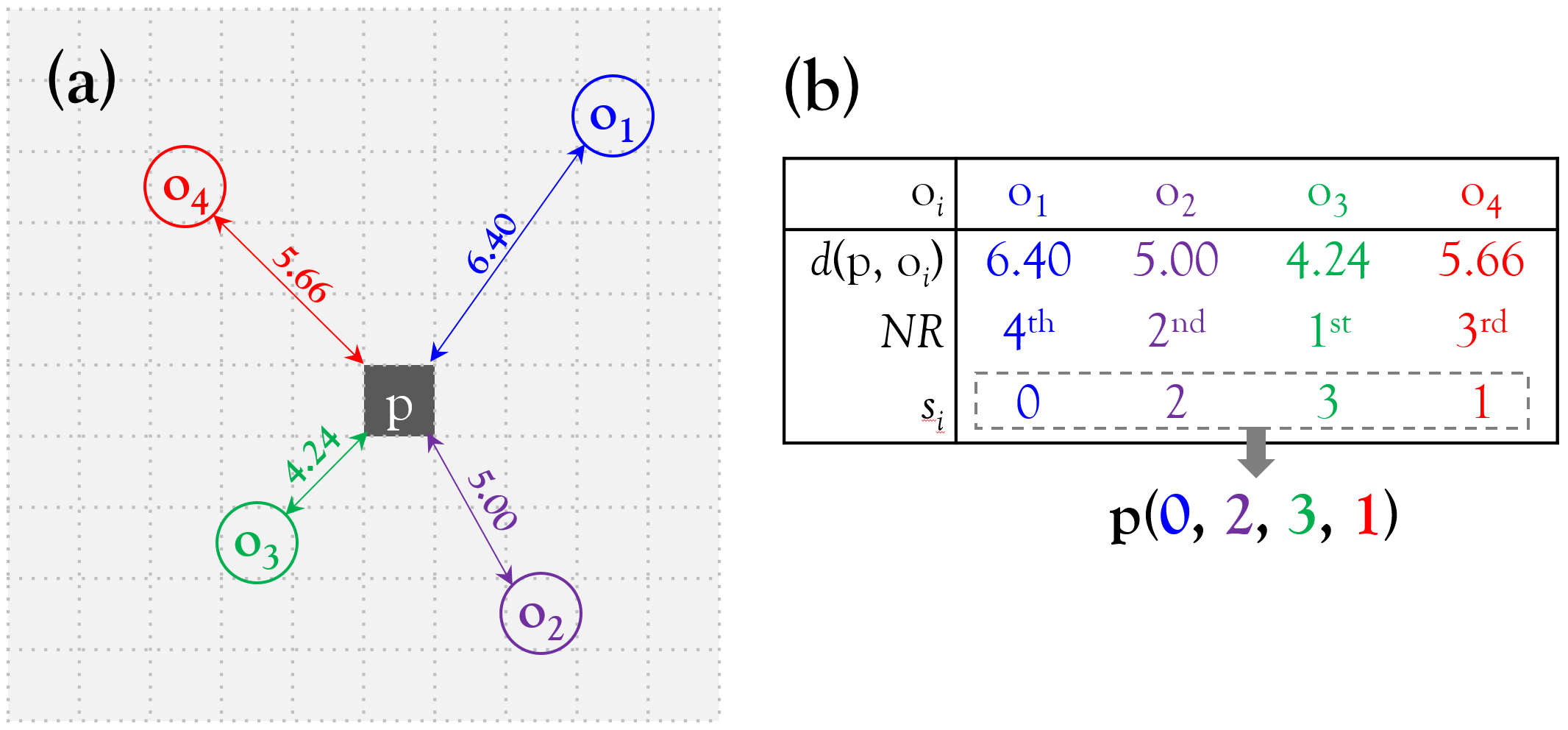}  \caption{The process of
generating a SCM from four objects in a discrete space by ranking distances
and merging points. (a) A discrete space ($10\times10$ pixels) containing 4
objects \{$o_{1}$, $o_{2}$, $o_{3}$, $o_{4}$\}, and an object occupies at
least one pixel; (b) the table to calculate the point $p$ 's chromatic code.
$d(p,o_{i})$: the distance between $p$ and $o_{i}$; $NR$: the nearest ranking
of $d(p,o_{i})$; For example, $NR=1$ means $o_{3}$ is the nearest object to
$p$; $s_{i}$: the $i$-th (or object $o_{i}$'s) subcode of $p$'s chromatic
code. $s_{i}=n-NR$, where $n=4$ is the number of the objects. Therefore, $p$'s
chromatic code is $(s_{1},s_{2},s_{3},s_{4})=(0,2,3,1)$. }%
\label{fig:3}%
\end{figure}

Beside the above bisector-dyeing method, there is also a nearest-ranking
method to generate a SCM that can be more conveniently used in a discrete
space (such as the digital images or raster model). Fig. \ref{fig:3} shows an example to
encode a point in a $10\times10$ raster space containing 4 objects. The
nearest-ranking method consists of the following steps.

\begin{enumerate}
\item Calculating distance. For any point $p$ in the space, calculating the
distance $d(p,o_{i})$ between $p$ and object $o_{i}$. In Fig. \ref{fig:3}(a), the
Euclidean distances between $p$ and $o_{1}$--$o_{4}$ are 6.40, 5.00, 4.24, and
5.66, respectively.

\item Sorting distance. Sorting the distance $d(p,o_{i})$ from the shortest to
the longest, and correspondingly ranking them as the 1$^{st}$, 2$^{nd}$,
3$^{rd}$, \ldots, and $n$-th nearest distances between $p$ and $o_{i}$.

\item Converting a rank into a subcode. Computing $s_{i}=n-rank$ for each
object $o_{i}$, where $n$ is the number of objects in the space. Note that a
rank is typically an ordinal number which cannot involve algebraic operations,
but in SCM, a rank is converted to an integer. For example, $rank(1^{\text{st}%
})=1$, and $rank(k^{\text{th}})=k$. Therefore, the distances (6.40, 5.00,
4.24, 5.66) is ranked as (4$^{\text{th}}$, 2$^{\text{nd}}$, 1$^{\text{st}}$,
3$^{\text{rd}}$), and since $n=4$, then $s_{i}=(n-4,n-2,n-1,n-3)=(0,2,3,1)$,
which is the chromatic code of $p$, see Fig. \ref{fig:3}(b).
\end{enumerate}

Following the above steps, the points with the same chromatic codes are merged
into a cell which also has the same chromatic code, and then the discrete
space is turned to a SCM space with many chromatic cells.

\section{CM algorithms for generating diverse VDs and new tessellations}

Cells in an SCM space can perform an important operation: merging multiple
cells to form a new subspace called a cluster. The chromatic code of a cluster
is simply the algebraic sum of the chromatic codes of all the merged cells.
For example, the merging of cells $c_{1}(2,3,1,0)$ and $c_{2}(3,2,1,0)$
generates a cluster $z\{c_{1},c_{2}\}$ with the chromatic code $(5,5,2,0)=(2,3,1,0)+(3,2,1,0)$.
When cells are merged, there are no restrictions on their number or location.
In practice, however, clusters are typically generated by cell merging rules
that apply only to cells with specific chromatic features, and these clusters
then collectively form novel spatial tessellations. The Voronoi region or
polygon of each object in a VD is actually a cluster, and each type of VD is a
unique spatial tessellation formed by using a specific merging rule to merge cells.

\begin{figure}[h]
\centering
\includegraphics[width=1.0\textwidth]{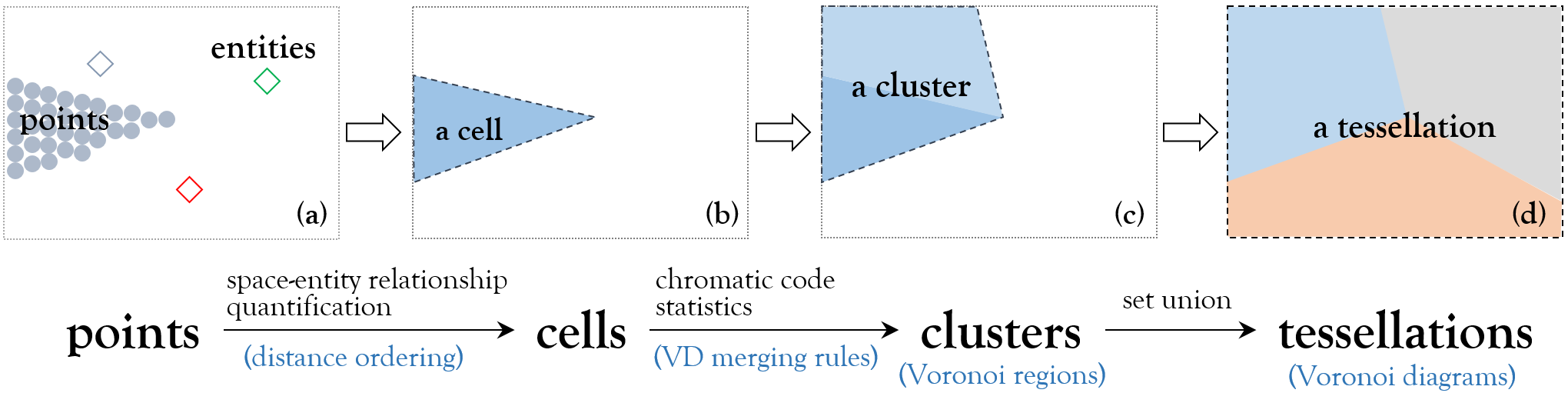}  \caption{The process of the
cell-merge algorithm to generate VDs in a discrete space. (a) Points with the same distance order in the space are merged into a cell in (b). Based on a merging rule, cells with the same or similar
statistical features of chromatic codes are merged into a cluster in (c),
i.e., a Voronoi region. The union of all clusters (Voronoi regions) generates
a new spatial tessellation (Voronoi diagram) in (d).}%
\label{fig:4}%
\end{figure}

Fig. \ref{fig:4} illustrates the process of generating VDs and other tessellations in
SCM. First, points with the same distance order are merged into a cell (Fig.
\ref{fig:4}(a)--(b)). Then, cells with the same statistical features of chromatic codes
are merged into a cluster, which is equivalent to a Voronoi region (Fig.
\ref{fig:4}(c)). Finally, all clusters together form a tessellation, such as a Voronoi
diagram (Fig. \ref{fig:4}(d)). The rules for merging cells into clusters are critical in
determining the resultant tessellations. The details of the CM algorithm are
given in the following six steps.

Step 1. Assign attributes to each point $p$ in space $\mathbf{S}$. Space
$\mathbf{S}$ refers to a theoretical or practical space for research or
applications, such as 2D or 3D Euclidean space, spherical or conical surfaces,
or network spaces (e.g., Network-VD). The points in $\mathbf{S}$ are organized
into a spatial point dataset, see Table \ref{tab:1}(a). The attributes $i$ and $j$ of
$p$ represent its spatial coordinates. In a 2D Euclidean space, vector
coordinates are denoted as $(x,y)$, while grid coordinates are represented by
row and column indices $(i,j)$. The $obj$ attribute of $p$ indicates whether $p$
belongs to an object: $obj=1$ means $p$ is a point of an object, while $obj=0$
means $p$ is an empty point in $\mathbf{S}$. The $obs$ attribute indicates
whether $p$ is an obstacle in an Obstacles-VD: $obs=0$ means no obstacle, and
$obs=2$ marks $p$ as the spatial location of the obstacle with the index 2.

\begin{table}[b]
\caption{Data tables in a pixel-based SCM database. (a) The table of spatial
pixels (containing $10\times10=100$ pixels); (b) The table of chromatic codes
of pixels in a given distance metric; (c) The table of chromatic codes of
cells which were merged from the 100 pixels with the same codes in Table 1(b);
(d) The table of pixels for merging cells in different three distance metrics
$d_{1}$, $d_{2}$, and $d_{3}$. (e) The cell table generated by the four objects. (f) The cluster table to form VDs and new tessellations.}
\label{tab:1}
\centering
\begin{tabular}
[c]{c}%
\includegraphics[width=1.0\textwidth]{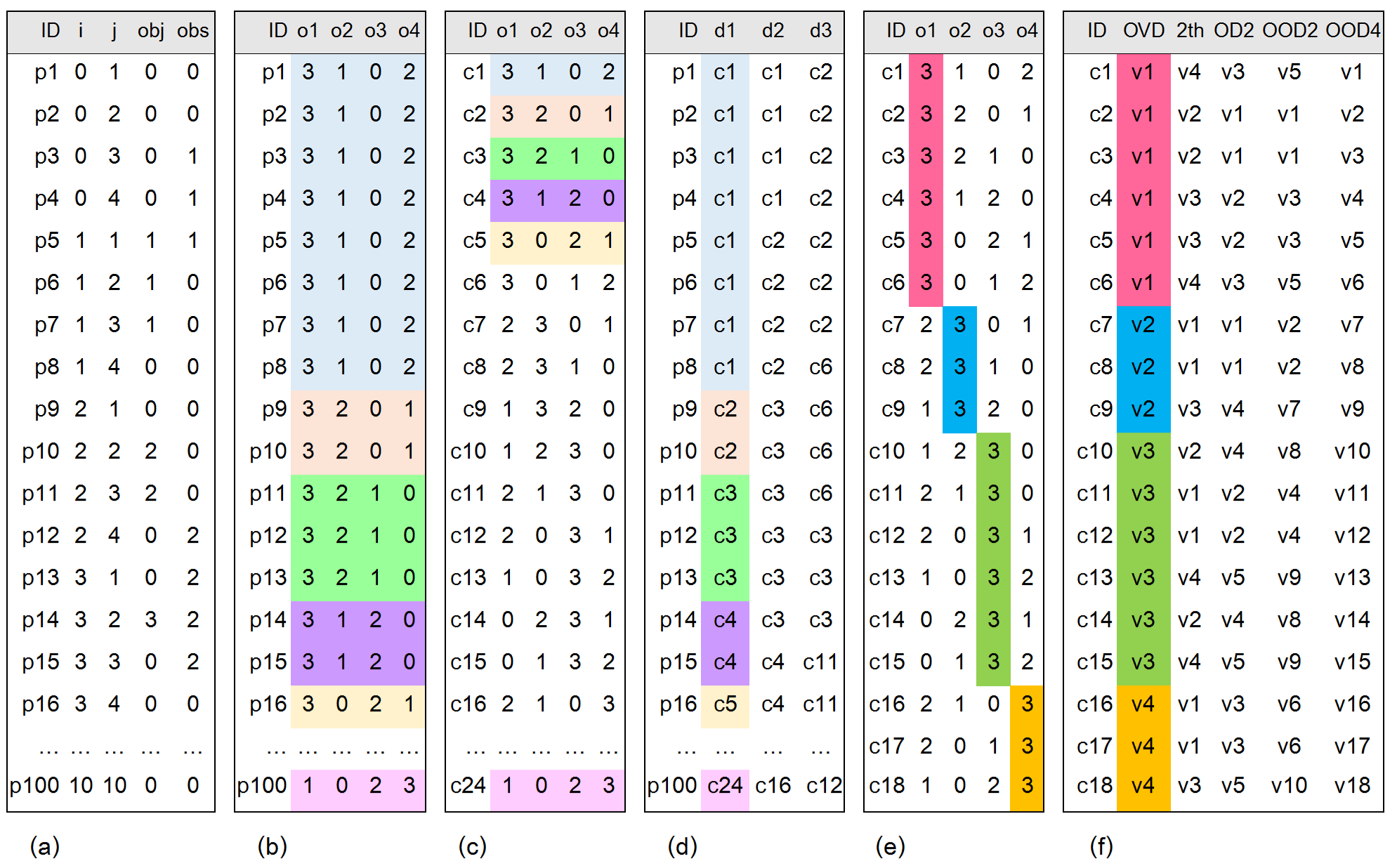}
\end{tabular}
\end{table}

Step 2. Using the nearest-ranking method to compute the chromatic code for a
point in the space. If multiple distances are equal, their corresponding
chromatic codes are assigned arbitrarily or according to predefined arrangements.

Step 3. Generate chromatic data tables for the SCM database. For all points
$p$ involved in spatial partitioning (sometimes excluding object and obstacle
points to reduce the computational load), repeat the computation and encoding
in Step 2. This process generates a chromatic code table for all spatial
points, see Table \ref{tab:1}(b). Based on the chromatic table of cells (see Table \ref{tab:1}(c),
which contains all possible chromatic codes of cells. For 4 objects, it
contains $4!=24$ types of chromatic codes of cells), each point $p$ is merged
into a specific cell $c$, forming a data table of point-to-cell merging, see
Table \ref{tab:1}(d). For example, points $p_{1}$--$p_{8}$ are all with the same
chromatic code $(3,1,0,2)$, and they are merged into the cell $c_{1}$ in Table
1(c). The columns $d_{1}$, $d_{2}$, and $d_{3}$ indicate that the results are
derived from different distance metrics. For example, in Table \ref{tab:1}(d), using the
Euclidean distance $d_{1}$, the point $p_{1}$ is merged into cell $c_{1}$ in
Table \ref{tab:1}(c), meaning its chromatic code is $(3,1,0,2)$, however, using
weighted distance $d_{3}$, it is merged into the cell $c_{2}$, so the
chromatic code of $p_{1}$ changes to $(3,2,0,1)$.

Step 4. Organize all the cells into a chromatic code table. In the Step 3, all
points were merged into cells. These cells are then organized into a chromatic
code table, see Table \ref{tab:1}(e), which contains 18 unique cells generated from four
objects in a 2D plane. Each cell has a unique chromatic code, and the
attributes $o_{1}$, $o_{2}$, $o_{3}$, and $o_{4}$ represent the chromatic
components of (or chromatic distances to) the four objects. Note that the
difference between Table \ref{tab:1}(c) and Table \ref{tab:1}(e) is that the cells in Table \ref{tab:1}(c)
have all 24 possible codes, whereas those in Table \ref{tab:1}(e) only have 18 codes
that actually emerge in the space $\mathbf{S}$.

Step 5. Search and merge cells into Voronoi regions. Based on the chromatic
table generated in Step 4, retrieve all cells that share the same chromatic
code features, which indicate the spatial point-object relationships, and
merge them into a Voronoi region. For example, perform the following database
query, using an SQL database where the table name is cell\_code\_table:

\texttt{\textcolor{blue}{SELECT}
Index\ \textcolor{blue}{FROM}\ cell\_code\_table\ \textcolor{blue}{WHERE}\ o1\ =\ 3}%

The above query returns the cell indices which identify cells whose chromatic
distance to object $o_{1}$ is 3, meaning that spatial points in these cells
have the shortest Euclidean distance to $o_{1}$. Consequently, these cells can
be merged into the nearest-neighbor region of $o_{1}$, i.e., the ordinary
Voronoi region of $o_{1}$.

\texttt{\textcolor{blue}{SELECT}\ Index\ \textcolor{blue}{FROM}\ cell\_code\_table\ \textcolor{blue}{WHERE}\ o1\ =\ 0}%

The above query returns the cell indices which identify the cells whose
chromatic distances to the object $o_{1}$ are 0, meaning spatial points in
these cells have the longest Euclidean distances to $o_{1}$, and hence these
cells can be merged into the furthest neighbor of $o_{1}$, namely, the
furthest Voronoi region of $o_{1}$.

\texttt{\textcolor{blue}{SELECT}\ Index\ \textcolor{blue}{FROM}\ cell\_code\_table\ \textcolor{blue}{WHERE}\ o1\ =\ 3\ \textcolor{red}{OR}\ o2\ =\ 2}%

\setlength{\parskip}{0pt}\texttt{\textcolor{blue}{SELECT}\ Index\ \textcolor{blue}{FROM}\ cell\_code\_table\ \textcolor{blue}{WHERE}\ o1\ =\ 3\ \textcolor{red}{AND}\ o2\ =\ 2}%

\setlength{\parskip}{3pt}The above two queries return the indices of cells
whose chromatic distance to object $o_{1}$ is 3 and/or to the object $o_{2}$
is 2, meaning spatial points in these cells have the shortest Euclidean
distance to $o_{1}$, and/or the second shortest Euclidean distance to $o_{2}$.
Therefore, these cells can be merged into the order-2 and ordered order-2
Voronoi regions associated with $o_{1}$ and $o_{2}$.

Step 6. Join all Voronoi regions into a Voronoi diagram. Repeat Step 5 for
each object or subset of objects to generate its respective Voronoi region.
The union of these regions forms the corresponding Voronoi diagram. For
example, an OVD is generated by repeating Step 5 for all objects:

\texttt{\textcolor{blue}{FROM}\ i\ =\ 1\ \textcolor{red}{to}\ n}

\setlength{\parskip}{0pt}\texttt{\ \ \textcolor{blue}{SELECT}\ Index\ \textcolor{blue}{FROM}\ cell\_code\_table\ \textcolor{blue}{WHERE}\ o(i)\ =\ 3}%

\texttt{\textcolor{blue}{END}}

\setlength{\parskip}{3pt} The database queries and their parameters, as shown
in Step 5 and Step 6, are called cell-merge rules for generating various
spatial tessellations. For example, in the cell-merge rule of OVD, cells with the same color, see Table \ref{tab:1}(e), were merged into the clusters $v_{1}$ -- $v_{4}$ in Table \ref{tab:1}(f), and then generated an OVD. 

\begin{figure}[h]
\centering
\includegraphics[width=1.0\textwidth]{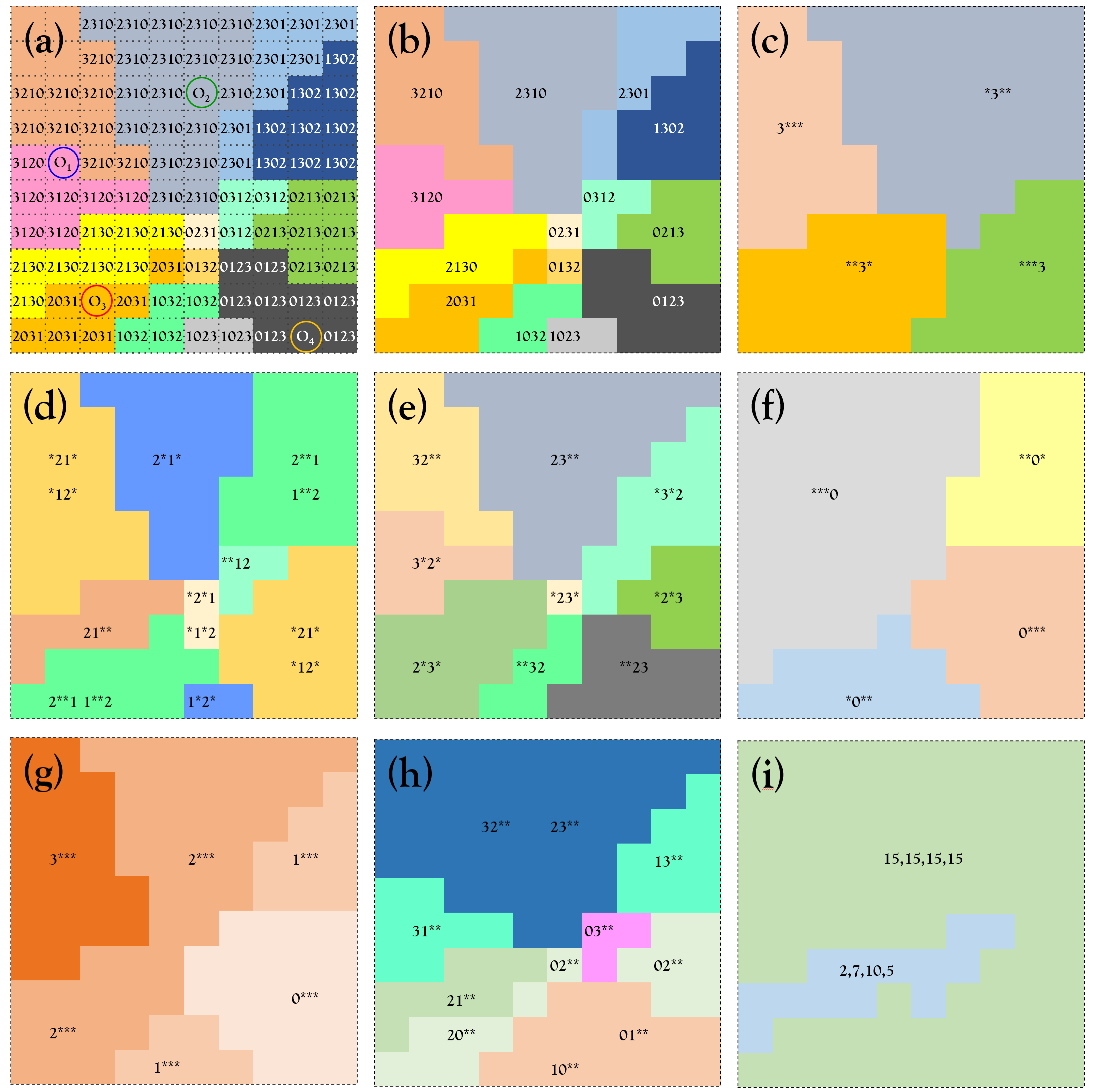}  \caption{The examples of
cell-merge algorithm to form spatial tessellations, based on a SCT generated
by 4 objects in a $10\times10$ grid space. (a) The four objects $o_{1}$ --
$o_{4}$ (the white circles), and each grid was encoded by a chromatic code;
(b) The cells by merging the grids with the same chromatic code; and further
merging cells using different rules to generate Voronoi diagrams: (c) OVD, (d)
2$^{\text{nd}}$ nearest VD, (e) 2$^{\text{nd}}$ ordered VD, (f) the furthest
VD, and new tessellations: (g) influence distribution, (h) competition level,
and (i) couple-cell diagrams. }%
\label{fig:5}%
\end{figure}

Fig. \ref{fig:5} shows the processes of point merging and cell merging in a $10\times10$
space for four objects (they locate at the four circles in Fig. \ref{fig:5}(a)). The
number in each spatial point (a pixel or a grid) is the chromatic code of that
point. Points with the same chromatic code have been merged into cells, as
shown in Fig. \ref{fig:5}(b). Each cell has an irregular shape as well as a unique
chromatic code.

When the cell-merge rule of OVD mentioned above in Steps 5 and 6 are applied, cells
with chromatic codes such as 3***, *3**, **3*, and ***3 (where * represents
any digit) are retrieved and merged to generate the OVD of the four objects,
as shown in Fig. \ref{fig:5}(c). When cells with chromatic codes such as *21*, 2*1*,
32**, 2**3, or ***0 are merged, they generate three types of VDs: the order-2, the ordered order-2, and the furthest VDs, respectively, as shown in
Fig. \ref{fig:5}(d)--(f).

The CM algorithm can generate new spatial tessellations different from VDs by
changing the cell-merge rules. In OVDs or other VDs, Voronoi regions are always
generated with respect to all objects, namely, for each object $p_{i}$ in
$\mathbf{P}$, the Eq. (\ref{Eq1}) returns a Voronoi region $v(p_{i})$. However, the
new tessellations can be generated with respect to one or more specific
objects rather than all of them. For the object $o_{1}$ in Fig. \ref{fig:5}(a), the
clusters in SCM can be formed by merging the cells with chromatic codes such
as 3***, 2***, 1***, and 0***, so the cell merging rule is as the following

\texttt{\textcolor{blue}{FROM}\ i\ =\ 0\ \textcolor{red}{to}\ n - 1}

\setlength{\parskip}{0pt}\texttt{\ \ \textcolor{blue}{SELECT}\ Index\ \textcolor{blue}{FROM}\ cell\_code\_table\ \textcolor{blue}{WHERE}\ o(1)\ =\ i}%

\texttt{\textcolor{blue}{END}}

\setlength{\parskip}{3pt} The new tessellation generated using the above rule
is called influence distribution diagram, as shown in Fig. \ref{fig:5}(g), which is similar to the buffer zones of the object $o_{1}$.

\begin{table}[b]
\caption{The merging-rule tables of some VDs and new tessellations in Fig. \ref{fig:5}. Tables 2(a)--(g) correspond to the diagrams Fig. \ref{fig:5}(c)
--(i), respectively.}%
\label{tab:2}
\centering
\begin{tabular}
[c]{c}%
\includegraphics[width=1.0\textwidth]{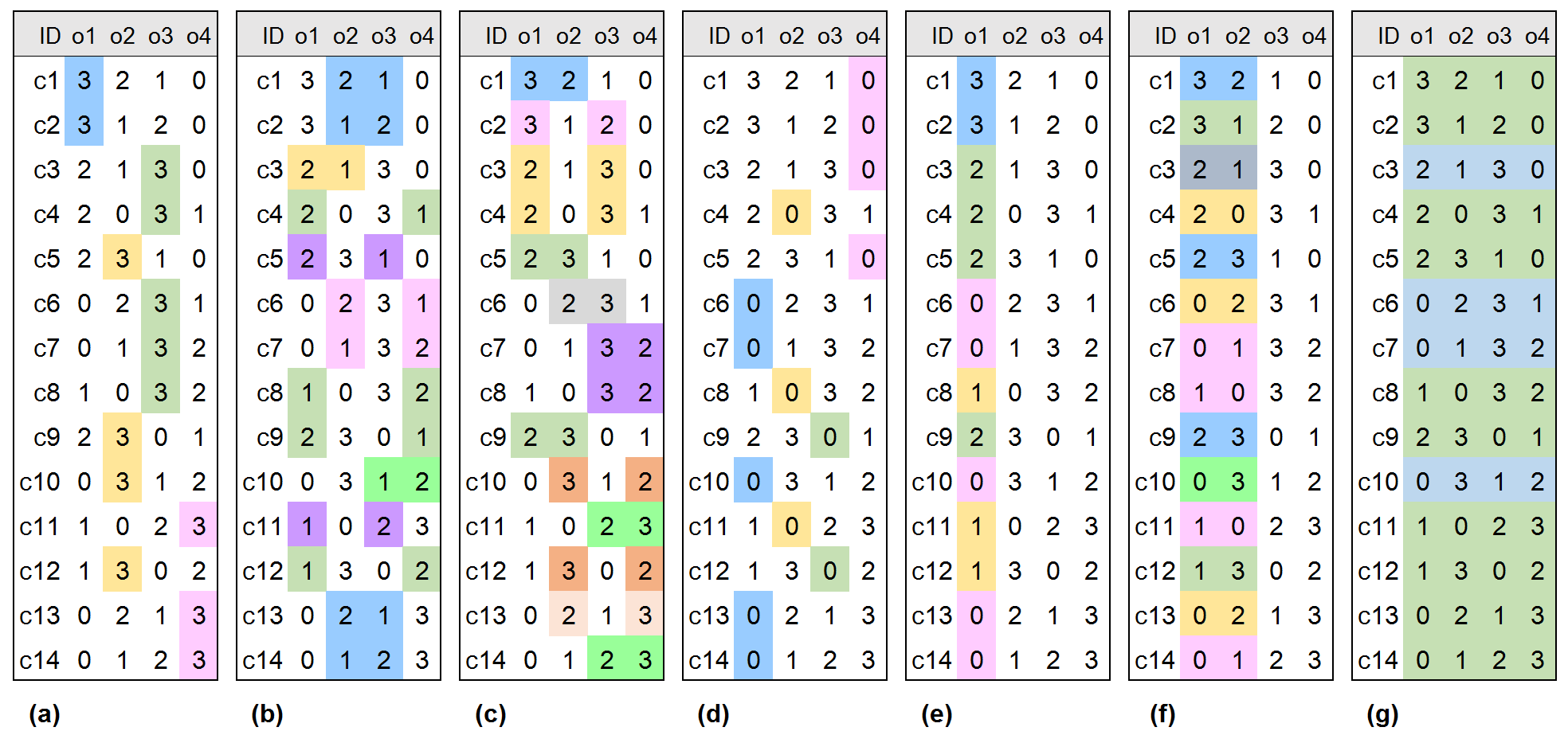}
\end{tabular}
\end{table}

Similarly, merging rules can be also with respect to two objects, for example,
$o_{1}$ and $o_{2}$, using the following rule:

\texttt{\textcolor{blue}{FROM}\ i\ =\ 0\ \textcolor{red}{to}\ n - 2}

\setlength{\parskip}{0pt}\texttt{\ \ \textcolor{blue}{FROM}\ j\ =\ i +
1\ \textcolor{red}{to}\ n - 1}

\texttt{\ \ \ \ \textcolor{blue}{SELECT}\ Index\ \textcolor{blue}{FROM}\ cell\_code\_table}%

\texttt{\ \ \ \ \ \ \textcolor{blue}{WHERE}\ (o(1)=i
\textcolor{red}{AND}\ o(2)=j) \textcolor{red}{OR}\ (o(1)=j
\textcolor{red}{AND}\ o(2)=i)}

\texttt{\ \ \textcolor{blue}{END}}

\texttt{\textcolor{blue}{END}}

\setlength{\parskip}{3pt} The new tessellation generated using the above rule
is called the competition intensity diagram, as shown in Fig. \ref{fig:5}(h).

The cell merging rule can also be based on the chromatic codes of the merged
clusters. For example, the couple-cell rule is defined as follows:

\begin{enumerate}
\item Define the couple-cell $c_{1}$ and $c_{2}$ such that $z\{c_{1}%
,c_{2}\}=(3,3,3,3)$.

\item Search all couple-cells from the cell\_code\_table and merge them into a
subspace called a coupled region.

\item Merge the cells without couples into a subspace called an orphaned region.

\item Form a new tessellation called the couple-cell diagram by taking the
union of the coupled and orphaned regions, see Fig. \ref{fig:5}(i).
\end{enumerate}

The above cell-merge rules are illustrated in the merging-rule tables shown
in Table \ref{tab:2}, where cells with the same color (sharing the same feature) are
merged into a cluster, and all clusters together generate a type of VD or a
new tessellation.

\section{Algorithm implementation and Discussion}

A software program called SCM Analysis was developed for spatial dyeing and
cell merging analysis. Its functions include adjusting spatial resolution,
setting objects via mouse clicks or file import, and modifying distance
metrics (e.g., using Minkowski distance and other weighting methods). To merge
cells, users specify different database queries and their parameters, and then
different groups of cells are identified and merged, generating various VDs as
well as other types of spatial tessellations.

\begin{figure}[h]
\centering
\includegraphics[width=1.0\textwidth]{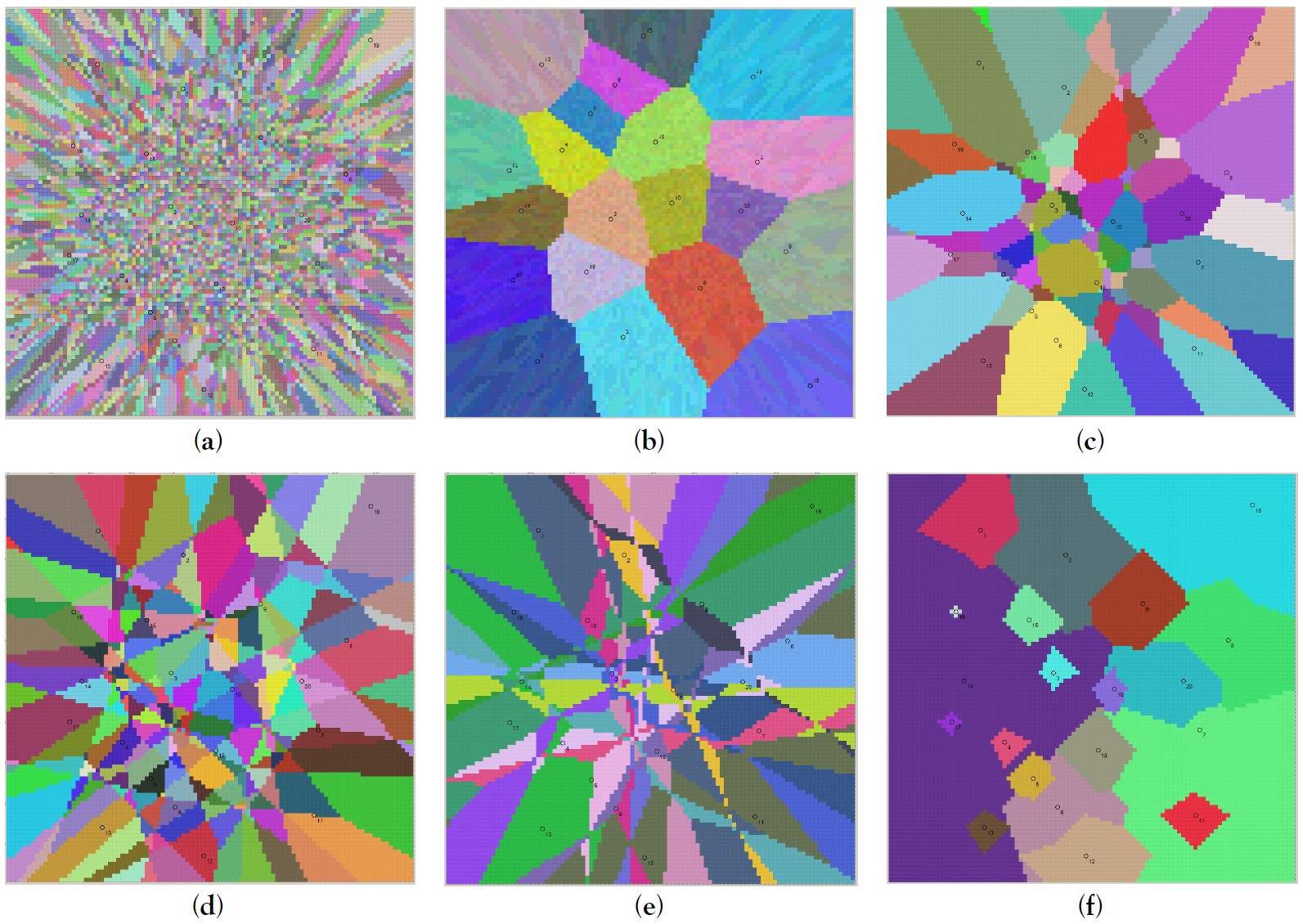}  \caption{Five types of VDs
generated from 20 objects in Euclidean distance: (a) the original cells in
SCM; (b) ordinary; (c) order-6; (c) ordered order-3; (d) 10-th nearest; and
(f) compoundly weighted in Manhattan distance.}%
\label{fig:6}%
\end{figure}

Fig. \ref{fig:6} illustrates the chromatic space and cells (Fig. \ref{fig:6}(a)) generated from 20
objects in a $100\times100$ pixel space. Fig. \ref{fig:6}(b)--(e) show the VDs obtained
by merging the cells in Fig. \ref{fig:6}(a) based on Euclidean distance and using
different cell-merge rules. The compoundly weighted VD in Fig. \ref{fig:6}(f) was merged
using the Manhattan distance metric. Such VDs can be rapidly generated by
updating the pixel indices of cells. For example, under the Euclidean metric
$d_{1}$, cell $c_{1}$ in Table \ref{tab:1}(d) contains eight pixels ($p_{1}$--$p_{8}$);
under the Manhattan metric $d_{2}$, these are updated to four pixels ($p_{1}%
$--$p_{4}$).

The algorithm tests and results presented above demonstrate that the CM
algorithm's advantages over traditional methods lie in its simplicity and
diversity. From a mathematical perspective, the various VDs generated from 20
objects represent different subset families of the set of all cells in Fig.
\ref{fig:6}(a). From an image-processing perspective, this corresponds to different
segmentations of the original image Fig. \ref{fig:6}(a). If a VD is analogous to a
building, the CM algorithm is like using bricks (cells) to construct various
buildings. In contrast, a traditional method of generating a VD is like
constructing one specific type of building; if we want a different type, we
must rebuild it using entirely new methods, without the ability to disassemble
and reuse the underlying bricks (cells).

In addition to generating various VDs, the CM algorithm can also generate new
spatial tessellations, such as the couple-cell diagrams, whose merging rules
are different from those of the VDs. If we treat various VDs as Gothic
architecture, then the CM algorithm can also use bricks to construct new
architecture in the Baroque style. A subcode $s_{i}$ can be treated as the
`quantized force' that object $o_{i}$ exerts on a cell, and a cell-merging rule
defines how these forces and their relationships are interpreted within a cell
or cluster. In OVDs, the force is a kind of gravity, and an ordinary Voronoi
region is a kind of gravitational field in which spatial points are attracted
to the object that has the greatest gravitational force (the largest subcode).
In influence distribution diagrams, the force is a kind of influence or
control power of an object, and the force decays as the subcode becomes lower.
In competition intensity diagrams, the force is a kind of tension. If two
forces are equally strong, then there will be intense competition between them
\cite{Zhu2006}. In couple-cell diagrams, the subcodes of the cluster formed by
couple-cells are all identical -- for instance, $(3,3,3,3)$ -- indicating that
the resultant forces among the four objects are balanced in a coupled region,
whereas in an orphaned region, the forces are not balanced.

Since SCM is the underlying spatial data model, the SCM-based CM algorithm can
be seamlessly combined with other SCM-based spatial analyses, such as spatial
topological analysis and spatial point pattern analysis \cite{Zhu2010,Zhu2015,Zhu2023}. These
SCM-based analyses and algorithms all study and explore patterns and features
in chromatic codes of cells and clusters, and can therefore be implemented
within a unified framework and with similar methods. The SCM functions as a
multifunctional tool (like a Swiss Army knife), applicable in many domains: VD
generation, spatial topology, spatial pattern recognition, etc. If traditional
methods were used, each domain would have its own analytical architecture,
strategies, and algorithms -- like using three separate tools (a screwdriver,
a wrench, and pliers) to solve a problem -- which would inevitably increase
the cost and complexity of spatial computation and analysis.

\section{Conclusion}

Based on the spatial chromatic model, a CM algorithm was proposed to generate
various Voronoi diagrams. Computer simulations and tests demonstrated that, by
searching the SCM database, the CM algorithm can rapidly generate a variety of
VDs such as OVD, weighted VD, ordered VD, and $k$-th nearest VD.

Compared with traditional VD algorithms, the advantage of the CM algorithm is
not in speed, but rather in its simplicity, uniformity, and diversity: (1)
There is no need to develop specific algorithms for each type of VD, so the CM
algorithm facilitates the use of multiple VDs for spatial analysis of objects
in practical applications; (2) The cell-merge rules for generating various VDs are
relatively simple and straightforward, requiring only changes in database
query variables and parameters. Since chromatic codes of cells are always
integers, and integers are preferred over floating-point numbers and strings
for sorting, recording, and computation, this further improves the efficiency
of database searching and processing; (3) In addition to generating various
well-known VDs, the CM algorithm can also generate new types of spatial
tessellations based on other cell-merge rules of chromatic codes, thus
providing more analytical tools for spatial applications; (4) The CM algorithm
can be integrated with other SCM-based spatial analysis tools, such as spatial
topology and point pattern recognition tools, so that these various tools can
be used within a unified framework.

If the spatial resolution is excessively high and the number of spatial
objects is very large, the CM algorithm may consume more computational power
and storage space during the initial construction of the SCM database.
However, with the continuous improvement of computer hardware capabilities
(e.g., CPU speed and disk storage capacity), these shortcomings are expected
to be largely overcome in the future.

\end{document}